\documentclass[]{aastex631}
\usepackage{xcolor}
\usepackage{enumitem}

\begin{document}

\title{An Alternative Explanation for the Helium Star Pulsar Binary J1928$+$1815: The Most Heavyweight Black Widow System to Date}

\author{Hang Gong}
\affiliation{Key Laboratory of Optical Astronomy, National Astronomical Observatories, Chinese Academy of Sciences, Beijing 100101, China}

\author{Alexey Bobrick}
\affil{School of Physics and Astronomy, Monash University, Clayton, Victoria 3800, Australia}
\affil{ARC Centre of Excellence for Gravitational Wave Discovery -- OzGrav, Australia}

\author{Francisco Garzón} 
\affiliation{Instituto de Astrofísica de Canarias, La Laguna, Spain. Departamento de Astrofísica, Universidad de La Laguna, Spain}

\author{Deven Bhakta}
\affiliation{Department of Astronomy, University of Virginia, 530 McCormick Road, Charlottesville, VA 22904, USA}
\affiliation{National Radio Astronomy Observatory, 520 Edgemont Road, Charlottesville, VA 22903, USA}

\author{Thomas Maccarone}
\affiliation{Department of Physics and Astronomy, Texas Tech University, Box 41051, Lubbock TX 79409-1051 USA}

\author{Sangita Kumari}
\affiliation{National Centre For Radio Astrophysics, Tata Institute of Fundamental Research, Pune 411007, India}
\affiliation{National Astronomical Observatory of Japan, 2-21-1 Osawa, Mitaka, Tokyo 181-8588, Japan}

\author{Nieves Castro Rodríguez}
\affil{GRANTECAN, Cuesta de San Jos\'e s/n, E-38712, Bre\~na Baja, La Palma, Spain}
\affil{Instituto de Astrof\'\i sica de Canarias, V\'\i a L\'actea s/n, E-38200 La Laguna, Tenerife, Spain}

\author{Antonio Cabrera-Lavers}
\affil{GRANTECAN, Cuesta de San Jos\'e s/n, E-38712, Bre\~na Baja, La Palma, Spain}
\affil{Instituto de Astrof\'\i sica de Canarias, V\'\i a L\'actea s/n, E-38200 La Laguna, Tenerife, Spain}

\author{Arash Bahramian}
\affiliation{International Centre for Radio Astronomy Research, Curtin University, GPO Box U1987, Perth, WA 6845, Australia}

\author{Jifeng Liu}
\affiliation{Key Laboratory of Optical Astronomy, National Astronomical Observatories, Chinese Academy of Sciences, Beijing 100101, China}
\affil{School of Astronomy and Space Science, University of Chinese Academy of Sciences, 100049, Beijing, China}
\affil{Institute for Frontiers in Astronomy and Astrophysics, Beijing Normal University, 102206, Beijing, China}

\correspondingauthor{ghang.naoc@gmail.com; alexey.bobrick@monash.edu; francisco.garzon@iac.es; thomas.maccarone@ttu.edu} 



\begin{abstract}
We present the results of deep near-infrared imaging of the recently discovered helium star pulsar binary J1928$+$1815 situated in the Galactic plane. Our observations did not achieve significant detections, providing limiting magnitudes of J=23.7 and H=22.2, which are both 2.4\,magnitudes deeper than the expected J and H magnitudes for a modeled stripped helium star with a mass of 1\,$\rm M_{\odot}$ after extinction. Although we cannot completely rule out the possibility of more significant extinction and the exact evolutionary status of the supposed helium star is uncertain, by comparing J1928$+$1815 with other pulsar binaries, we propose a natural alternative solution: that J1928$+$1815 is a heavyweight black widow system with a massive ablated white dwarf. Due to the pulsar's relatively high spin-down power and short orbital separation, the irradiation heating timescale is uniquely shorter than the cooling timescale for the WD companion. As a result, the WD effectively boils, with its outer layers expanding, overfilling the Roche lobe and producing low-density binary-scale haze opaque in the radio band. If this interpretation is correct, J1928$+$1815 would represent a new category distinct from canonical lightweight black widow systems. Radio eclipses can occur in pulsar binaries across a wider range of WD companion masses than previously thought. Therefore, they do not serve as a definitive indicator of a helium star without its direct detection. 
We contend that a spectroscopic identification remains the smoking gun for its existence. Given the crowding in this field, an \textit{HST} imaging in the near-infrared band would provide even better constraints. 

\end{abstract}



\section{Introduction} 
Millisecond pulsar (MSP) binaries with a massive white dwarf (WD) companion on a circular orbit are believed to have transitioned through a helium star (He star) stage in their evolutionary history \citep{2002MNRAS.331.1027D,2012MNRAS.425.1601T, 2018A&A...619A..53T}. In one scenario, called `direct NSWD' in \citet{2018A&A...619A..53T}, the He star forms when the neutron star (NS) strips the hydrogen envelope from the non-degenerate companion in a high-mass X-ray binary (HMXB) through a short common envelope (CE) phase. In the resulting short-period circular He Star-NS binary, the He star later expands due to the exhaustion of its helium core, entering a helium giant phase and starting a second mass transfer phase known as Case BB Roche-lobe overflow (RLOF), spinning up the NS, and eventually evolving into a massive WD-MSP binary. Alternatively, the NS may form when the secondary is already a He Star (`reversed NSWD' scenario in \citealt{2018A&A...619A..53T}). In this case, the He star-NS binary is eccentric because of the natal kick acquired by the newborn NS. The following phase of interaction with the evolved He Star circularizes the binary and spins up the NS, also producing the MSP-WD binary. In both cases, the He star stage serves as a bridge between massive main sequence progenitors and compact binary pulsars \citep[for further details, see ][]{2023pbse.book.....T}. However, in both of the scenarios above, it is relatively challenging to produce a He Star-NS binary that is both circular and contains a MSP with a $1$~--~$10\,\text{ms}$ spin period. In the `direct NSWD' channel, the He Star-NS binary is naturally circular, but the NS must be spun up during the preceding HMXB or common envelope phase. Currently known detached post-common envelope NS candidates typically have spin periods between a few times $10\,\text{ms}$ and $100\,\text{ms}$ \citep{2005MNRAS.363L..71D}, which are large compared to the $1$~--~$10\,\text{ms}$ spin period typically observed in MSPs \citep{2017JApA...38...42M}. In the `reversed NSWD' channel, the He Star+NS binary is both eccentric and contains a pulsar with a lower spin period.

A very recent discovery of J1928$+$1815 \citep{2025Sci...388..859Y} (hereafter YH2025) claims to have captured the He star stage in MSP binaries for the first time, providing evidence for the evolutionary theory. The system was initially discovered to have a spin period of 10.5\,ms and a dispersion measure (DM) of $\approx346\rm{\,pc\,cm^{-3}}$ by the FAST Galactic Plane Pulsar Snapshot survey \citep[gpps0121 in][]{2021RAA....21..107H}. Following over three years of timing observation in the L band, J1928$+$1815 was identified to have an orbital period of 3.60\,hours. Based on the determined mass function, the companion's minimum mass (i=$90^\circ$) is 1.15\,$\rm M_{\odot}$ for a typical pulsar mass of 1.4\,$\rm M_{\odot}$, or 0.97\,$\rm M_{\odot}$ for a less likely pulsar mass of 1\,$\rm M_{\odot}$ \citep{2015ApJ...812..143M,2022NatAs...6.1444D,2025PhRvL.134g1403M}. In contrast to other pulsar systems within this companion mass range, J1928$+$1815 uniquely exhibits a regular radio eclipse that lasts for 17\% of its orbital period. 
YH2025 argues that J1928$+$1815's companion cannot be a main sequence star because such a star lacks a sufficiently compact mass-radius relation to avoid Roche lobe overflow. However, the companion must possess both a stellar size and outflow material (stellar wind) in a highly inclined orbit, which suggests that a helium star is a plausible alternative. The companion upper mass limit, set at 1.6\,$\rm M_{\odot}$, is determined based on the mass-luminosity relation in \citet{2023ApJ...959..125G} and its non-detection in the Galactic Plane Survey (GPS\footnote{\url{http://www.ukidss.org/surveys/gps/gps.html}}) conducted by the UK Infrared Telescope \citep{2007MNRAS.379.1599L}. Using the stellar evolution code MESA\citep{2011ApJS..192....3P,2019ApJS..243...10P}, simulations (Figure S8 of YH2025) adopting a helium star mass of 1.15\,$\rm M_{\odot}$, a pulsar mass of 1.4\,$\rm M_{\odot}$, and an orbital period of 3.6\,hrs were performed. The results demonstrate the system's stability for 10$^7$ years prior to the helium star's expansion, a time scale that is not negligible, thereby increasing the likelihood of the companion being a helium star.

On the other hand, the vast majority of pulsar binary radio eclipses are observed near the lower mass end of the companion spectrum, commonly referred to as spider pulsars. These include degenerate black widow (BW) systems and non-degenerate redback (RB) systems \citep{2013IAUS..291..127R}. The former MSP binaries typically have orbital periods less than one day and BW companion masses significantly less than 0.1\,$\rm M_{\odot}$, which are much lower than the companion masses in RB pulsars. They are believed to be descendants of low-mass X-ray binaries and progenitors of isolated millisecond pulsars due to the process of ablation \citep{1988Natur.334..227V,2013ApJ...775...27C}. Typically, the companions of BWs have radius of 0.1\,$\rm R_{\odot}$ \citep{2019ApJ...883..108D}.

Based on our near-infrared (NIR) imaging observations, and since it is challenging for He Star-NS binaries to contain an MSP on a circular orbit, we propose an alternative solution for J1928$+$1815. If the current estimates of extinction and the properties of the He star are correct, we can exclude a significant portion of the parameter space associated with the He star scenario. Consequently, we consider a massive ablated WD as a possible and complementary solution, classifying J1928$+$1815 as a heavyweight WD BW system distinct from the canonical light degenerate ones. 

\section{EMIR@GTC Observations and Data Analysis}
YH2025 has conducted a meticulous calculation indicating that, in the optical band, in the direction of J1928$+$1815, ground-based telescopes can barely detect a 1--1.6\,$\rm M_{\odot}$ He star at a distance of 8\,kpc, representing a moderate choice between the two DM-based distances of 7.2 and 9.7\,kpc. However, in the NIR band, which is considerably less affected by the significant extinction in the Galactic plane, an 80-second exposure in the {\it J} or {\it H} band conducted with the 3.8\,m-class UKIRT has either approached or exceeded part of the expected magnitudes of the He star.  

Aiming at detecting J1928$+$1815 in the NIR, a deep imaging program (Program ID: GTC04-24ADDT; PI: Gong) was conducted using the EMIR (Espectrógrafo Multiobjeto Infra-Rojo) instrument \citep{2022A&A...667A.107G} with {\it J} and {\it H} filters on the nights of May 17 and 19, 2024, at the 10.4\,m Gran Telescopio Canarias (GTC) in Spain. EMIR is a wide-field cryogenic camera-spectrograph that operates in the range of 0.9--2.5\,$\mathrm{\mu m}$. Observations were taken using a four-point dither pattern, with pointings distributed at the center and along the perimeter of a circle with a $10''$ radius. In the {\it J} band, two cycles were obtained per night, with two 30-second exposures at each pointing. In the {\it H} band, five cycles were obtained per night, with four 20-second exposures per pointing. The elevation angle was above 60$^\circ$ with seeing conditions of $1''$ on the first night and $0.8''$ on the second night.  

The images were processed using PyEmir \citep{2010ASPC..434..353P}. Given the challenge of detecting a very faint source against a high sky background, particular care was taken in the reduction of the observations, the details of which are provided below. At first, a visual inspection of all image frames was conducted to identify any spurious features that would necessitate frame removal; however, no such features were identified.

Using PyEmir\footnote{\url{https://pyemir.readthedocs.io/en/stable/user/highlevel.html}}, the following steps were performed:

\begin{itemize}[noitemsep, topsep=0pt, parsep=0pt, partopsep=0pt]
    \item A bad pixel mask was applied, and the individual frames were reprojected to correct for geometric distortions. Flat-field correction was not applied at this stage. Each group of images within the same dither pointing was co-added.
    \item A superflat was derived and applied. This was obtained from the full set of co-added images, exploiting the dithered nature of the data. The superflat was constructed by computing sigma-clipped median values along the frame sequence for each pixel. It was subsequently normalised before application to each co-added image.
    \item Sky background correction was determined and applied on a per-image basis, using a predefined number of images taken before and after the frame to be corrected. This process began with realigning the images from each night, using World Coordinate System (WCS) information in conjunction with 2D cross-correlation of sub-images centered on bright targets. Point-like objects were then detected and masked in the image subset used for sky correction. The sky signal at each pixel was estimated as the scale-weighted median of the corresponding pixel in the selected images. The resulting sky frame was scaled to match the median value of the frame to be corrected and subsequently subtracted.
\end{itemize}

Finally, offsets between the corrected co-added images were refined after cross-correlation with {\it Gaia DR3} \citep{2016A&A...595A...1G, 2023A&A...674A...1G}, and the complete set for each night was combined. The final combined image per night was cross-correlated with {\it Gaia DR3} to refine the WCS, and both nights were subsequently merged. The original astrometry from the GTC is of the order of $0.5''$ in accuracy, which may be sufficient for other projects. However, in this case, given the high density of objects on each frame, it is not. Following correlation with {\it Gaia}, we have obtained mean values for accuracy of $0.06''$, measured as differences between the astrometry in the image frames and in {\it Gaia}. The fully combined image was then cross-correlated with {\it 2MASS} \citep{2006AJ....131.1163S} to determine the photometric zero-point of the observations. The {\it 2MASS} sources used for photometry were selected by excluding those with neighbouring objects, measured using {\it SExtractor} \citep{refId0}, within $5''$ and with a flux greater than 50 counts in a circular aperture of 3 pixels radius. To measure the flux of the objects in the EMIR frames, the total flux within a circular aperture of a given radius was measured, and the local background to be subtracted was determined as the median value in a 2-pixel-wide ring surrounding the aperture. Count values in the background were excluded by a 3-sigma clipping process. 

In the final image, photometry of the target object was performed in the same way as in the case of {\it 2MASS} photometry, using a circular aperture with a diameter of $1.2\times$\textit{FWHM$_{seeing}$}, where the \textit{FWHM} was measured from isolated, moderately bright sources within the same image. Aperture diameters of 6 and 8 pixels, pixel size of $0.1945''$, were tested, yielding no significant differences in the results. Care was taken to ensure that neither the aperture nor the background annulus overlapped with nearby sources of moderate brightness, in addition to the sigma--clipping process.

\begin{figure}
\centering
\includegraphics[width=.48\textwidth, trim= 0 0 0 45,clip]{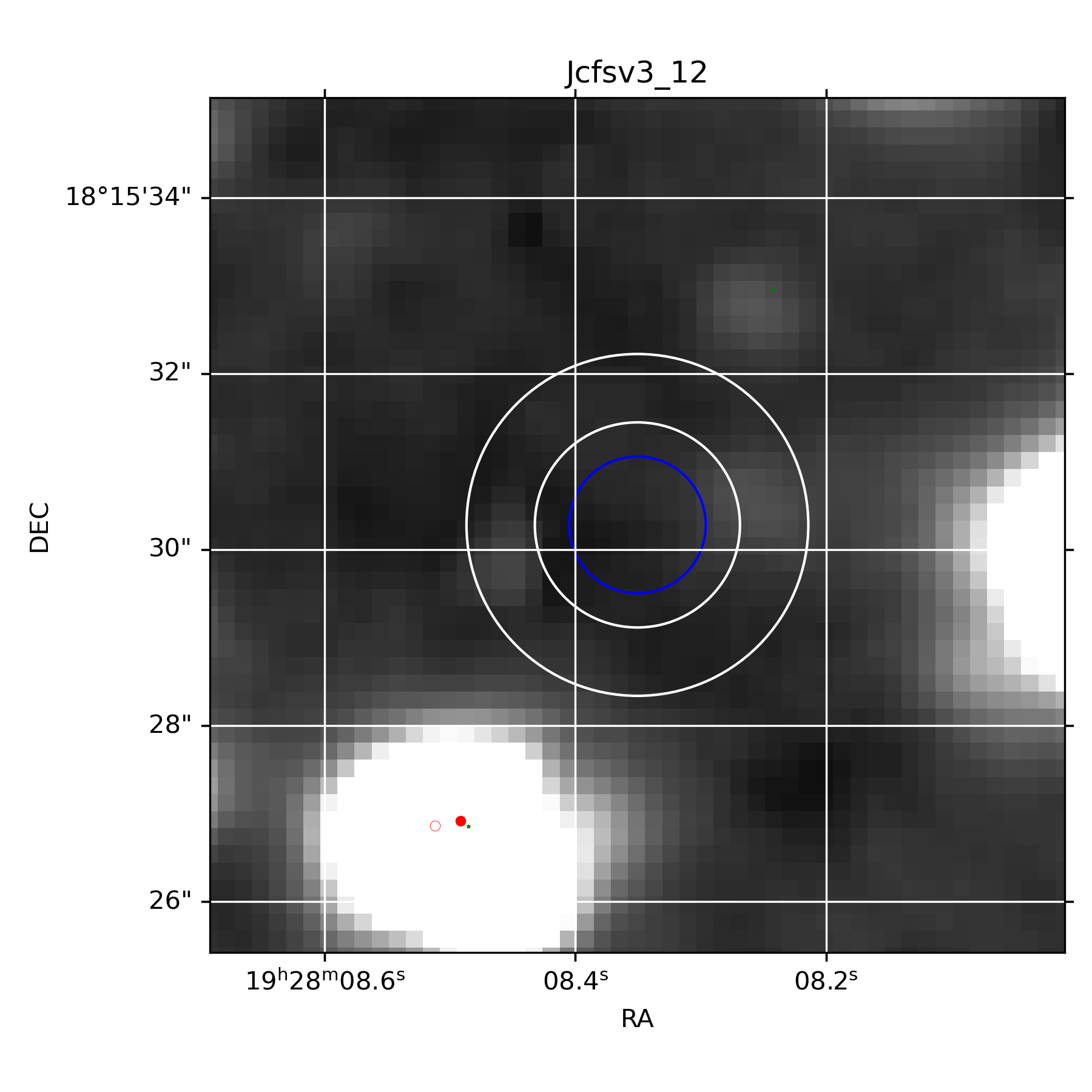}
\includegraphics[width=.48\textwidth, trim= 0 0 0 45,clip]{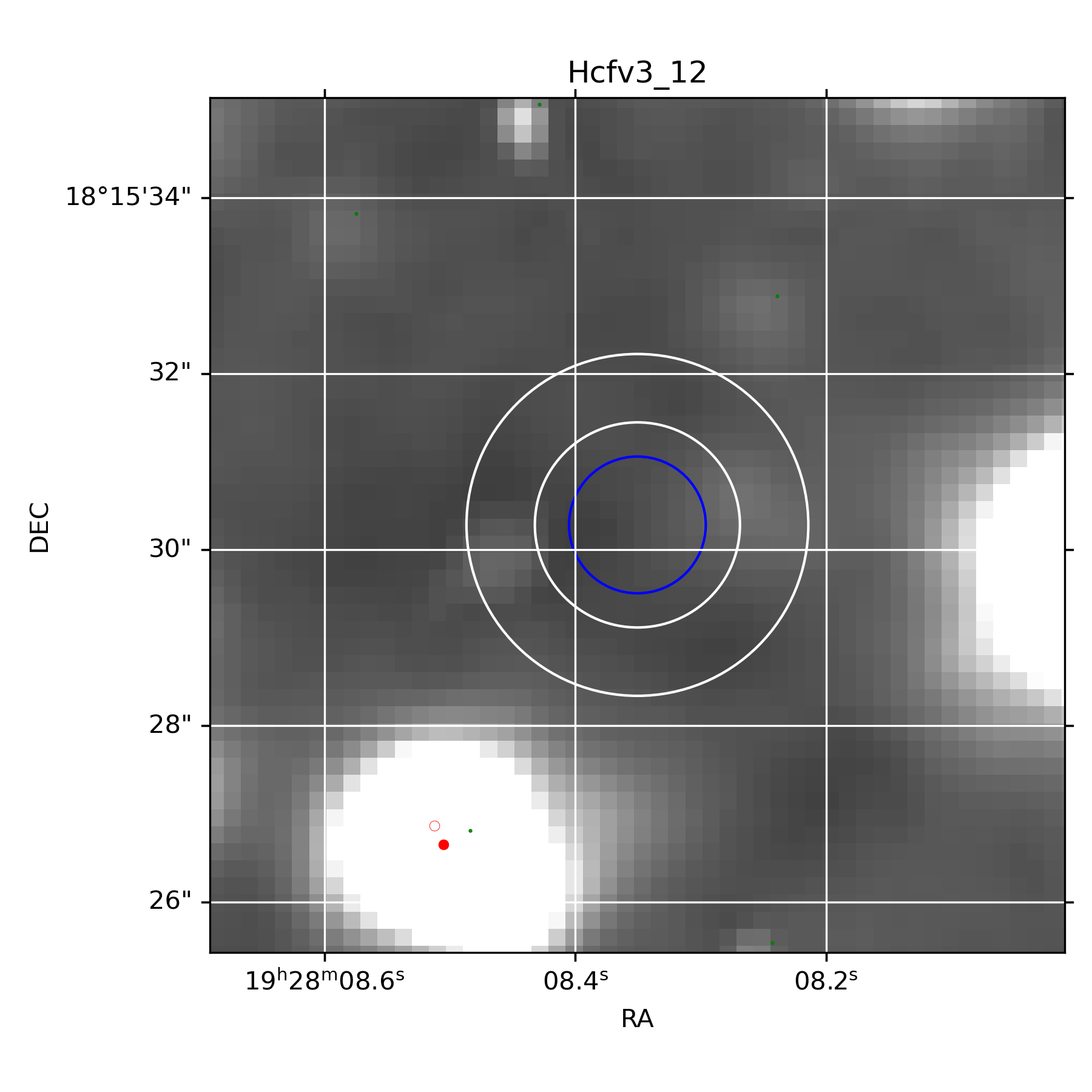}

\caption{Zoom in the {\it J}, left, and  {\it H}, right, final frames around the position of J1928$+$1815, which is in the center of the blue circle. The annulus in white surrounding the aperture sets the area to measure local background. To illustrate the astrometric accuracy of the analysis, the open and filled red
circles and the green point in the bright source on the lower left represent the positions of that object in {\it Gaia}, EMIR frame {\it WCS} and the centroid measured by {\it SExtractor}, respectively.} 
\label{fig_emir}
\end{figure}

Following reduction, no signal from the object was observed, as can be seen in Figure~\ref{fig_emir}. However, we can provide an estimated lower magnitude limit of brightness at 5-sigma of the local noise. 
The results yield Vega magnitudes of $23.728\pm0.150$ for {\it J} and $22.207\pm0.276$ for {\it H}, respectively.

\section{Gamma Ray Analysis}
Due to the substantial spin-down power in the binary, we performed a search for gamma-ray pulsations from J1928$+$1815 in the \textit{Fermi}-LAT photon data. We selected Pass 8 {\tt SOURCE}-class photons \citep{2013arXiv1303.3514A} detected by the \textit{Fermi}-LAT between 2008 August 5 and 2024 August 13 with an energy range of 0.1-5\,GeV photons, from within a 5\,deg radius around the radio position of the pulsar. Photon weights (i.e. the probability of each photon being associated with the pulsar) are vital for gamma-ray pulsation searching and timing. These weights were calculated using the {\tt gtsrcprob} routine within {\tt Fermitools} and the {\tt P8R3\_SOURCE\_V3} instrument response functions. The routine derives probabilities for the photons using a combination of a spectral model (acquired via the \textit{Fermi}-LAT 14-year Source Catalog \citep[4FGL-DR4;][]{2022ApJS..260...53A,2023arXiv230712546B} and a spatial model to distinguish events associated with the pulsar as opposed to a nearby point source \citep{2011ApJ...732...38K,2019A&A...622A.108B}.  

For the pulsation search, we used {\tt event\_optimize} within {\tt PINT}  \citep{2021ApJ...911...45L,2024ApJ...971..150S} along with the FAST timing solution. {\tt event\_optimize} \ uses Markov Chain Monte Carlo (MCMC) and Bayesian analysis to calculate the best values of the timing model parameters based on optimizing the model's likelihood. Due to the lack of a known \textit{Fermi} point source near the position of the pulsar, the chance of detecting pulsations was low. We searched for pulsations over the time range corresponding to the radio timing solution before attempting to extend the solution to cover the full 16-year \textit{Fermi}-LAT dataset. However, we were unsuccessful in detecting any pulsed signature in the events corresponding to the radio observations, which limited our ability to extend this to the full dataset.

\section{Discussion}

\begin{figure}
\centering
\includegraphics[width=.6\textwidth]{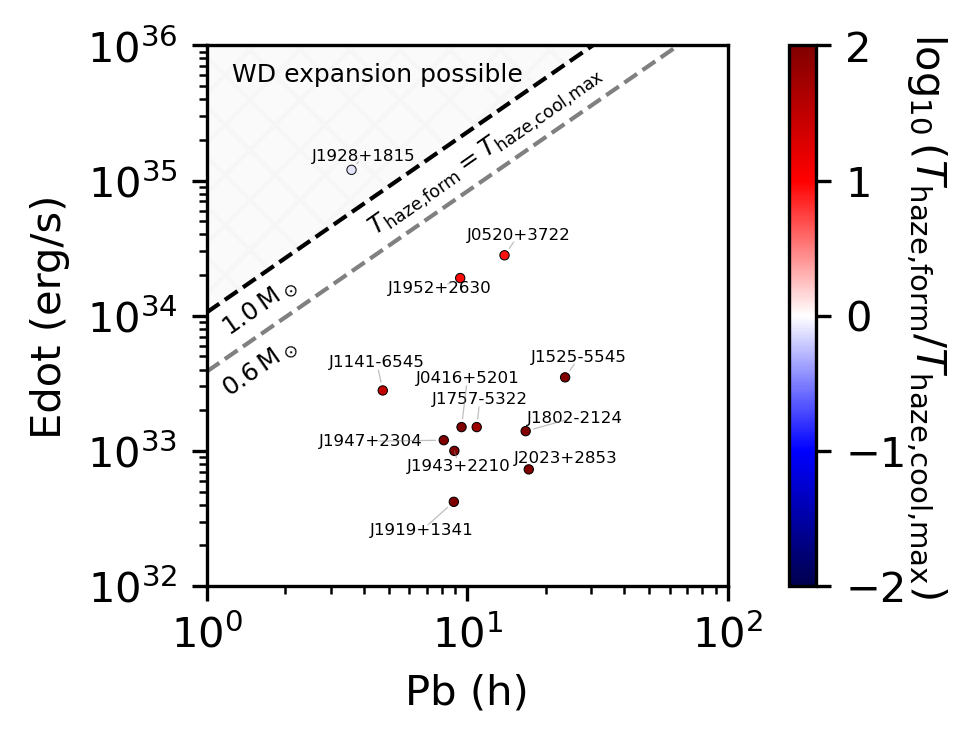}
\caption{The distribution of pulsars in Table~1 on the $\dot{E}-P_{\text{b}}$ plane. The colour map shows the ratio of heating timescale, $T_{\text{haze,form}}$, and the longest possible cooling timescale, $T_{\text{haze,cool,max}}$ (which is largest value for $T_{\text{haze,cool}}$ in Table~2), on a log scale. One may see that J1928$+$1815 is unique in having its pulsar spin-down power and orbital separation prevent it from cooling efficiently. We highlight the region facilitating the process, assuming a system with a $1\,\text{M}_\odot$ WD and a $1.4\,\text{M}_\odot$ NS, and show the boundary of the region with a black dashed line. For comparison, we show the same boundary for the case of $0.6\,\text{M}_\odot$ WD with a grey dashed line. Newly discovered pulsars in the highlighted region are likely to also exhibit similar phenomena.} 
\label{fig_psr_cmp}
\end{figure} 

The new observational limits are $3$ to $4$ magnitudes deeper than previously available (Table~S2 of YH2025). Accounting for extinction, our observations are 2.4 magnitudes deeper than the expected {\it J} and {\it H} band magnitudes for modeled stripped He stars and 3.1 magnitudes deeper than observed He stars with a mass of 1\,$\rm M_{\odot}$. Even for the extreme case of modelled naked He star, our limits are 1.6 magnitudes deeper. YH2025 used the mass-luminosity relation $\log(L_{\text{He}}/L_\odot)=3.45^{+1.05}_{-0.98}\log(M_{\text{He}/M_\odot})+2.41^{+0.75}_{-0.77}$. Extending this relation to lower masses or using the tables from \citet{2018A&A...615A..78G} implies in both cases $M_{\text{He}}=0.5\,\text{M}_\odot$. Therefore, assuming the originally estimated extinction from YH2025 is correct, the derived optical limits are well below the He star companion mass inferred from the pulsar timing measurements. 

In YH2025, the non-detection of J1928$+$1815 in the GPS is primarily attributed to the limited observation depth and significant extinction, which is already considerable but potentially may be even greater, in addition to uncertainties in mass and evolutionary status of the helium star. For instance, the presence of two spiral arms may enhance extinction depending on their precise relative positions to the pulsar system. Additionally, J1928$+$1815's DM-based distance may have a 30\% uncertainty. Furthermore, its uncertain local environment may present another source of error. These factors remain significant obstacles for our new NIR observations, making it impossible to completely eliminate their influence. Consequently, while our observations can exclude a broader parameter space, we do not wish to dismiss the helium star scenario; rather, we contend that a spectroscopic identification remains the smoking gun for their existence. However, from both observational and statistical perspectives, if the current evolutionary theory is correct, we speculate that a selection effect might be an initial misidentification of these systems as redback pulsar binaries, irrespective of the presence of eclipses, with subsequent spectroscopic evidence demonstrating that the companions are helium stars. Reliance on this single discovery channel (i.e., eclipsing) without directly seeing helium stars could instead undermine this scenario. Therefore, we propose a natural alternative explanation: J1928$+$1815 is a heavyweight BW system, representing a compact pulsar binary with the most massive irradiated WD companion to date. The distinction from canonical lightweight BW systems is that J1928$+$1815 possesses a significantly more massive donor.
 
The primary reason J1928$+$1815 is classified as a helium star pulsar binary in YH2025, distinguishing it from other compact radio pulsars with massive WD companions, is its extended radio eclipse. Table~1 presents the properties of J1928$+$1815 and all known pulsar binaries in the ATNF catalog \citep{2005AJ....129.1993M} with massive WD companions, where $\rm{P_b}$$<$1\,day and MinMass$>$0.5\,$\text{M}_{\odot}$. Among these, J1141$-$6545 is considered to follow a different evolutionary path, with its WD companion having formed earlier than the neutron star \citep{2000ApJ...543..321K}. The last ten systems have progressed beyond the He star phase and entered the WD stage, yet they exhibit binary parameters similar to J1928$+$1815. While a one-solar-mass He star is approximately 20 to 60 times larger than a massive WD \citep{2021Natur.595...39C,2005A&A...441..689A}, direct occultations by the companion star itself may play a minor role in generating the eclipses. Firstly, we show that J1928$+$1815 is special since, for its parameters, ablation due to pulsar spin-down energy causes the WD to expand on timescales shorter than thermal cooling timescales of the WD. In this case, the ablated low-density material surrounding the binary is opaque in the radio band and might thus be the reason for the observed radio eclipses \citep{1994ApJ...422..304T,2020MNRAS.494.2948P}. To quantify this scenario, we calculate the WD ablation rate, following \citet{1992MNRAS.254P..19S} as:
\begin{equation}
    \dot{M}_{\text{abl}}=0.5 f \frac{\dot{E}}{v_{\text{esc}}^2}\left(\frac{R_{\text{WD}}}{a_{\text{bin}}}\right)^2,
\end{equation}
where $v_{\text{esc}}$ is the WD surface escape speed, $R_{\text{WD}}$ is the WD radius based on mass-radius relation from \citet{2017MNRAS.467.3556B} and its estimated minimal mass, $a_{\text{bin}}$ is the binary orbital separation, which we take to be equal to $a_1(1+M_{\text{NS}}/M_{\text{comp}})$. Note that the mass-radius relation for white dwarfs is anti-correlated, unlike that of main-sequence and helium stars. Finally, parameter $f=0.2$ is a conservative estimate for the efficiency with which the pulsar irradiation energy can unbind the material. The mass-radius relation from \citet{2017MNRAS.467.3556B} is constructed based on a grid of spherically-symmetric stellar structure models of WDs based on a realistic Helmholtz equation of state for partially-degenerate matter \citep{2000ApJS..126..501T} and is consistent with existing analytic approximations, e.g. \citep{1988ApJ...332..193V}. We also note that the calculated binary semi-major axis is a good approximation since the orbit is circular and nearly edge-on, and the equality becomes exact for edge-on orbits. Applying this calculation to the pulsars in Table~1 and summarizing the results in Table~2, we see that the suggested WD in J1928$+$1815 has the highest ablation rate among known pulsars due to the large spin-down power and relatively close orbital separation. 

We further note that relatively little material is needed to make it radio opaque. Using Equation~16 from \citet{2017MNRAS.467.3556B}, which is based on radiative transfer equations, assuming Thompson cross-section for the fully ionized haze and an exponentially decreasing radial density profile on the scale of $R_{\text{haze}}$ (i.e. $\rho_{\text{haze}}(r)\propto \exp(-r/R_{\text{haze}})$), its optical thickness from the center outwards is given by:
\begin{equation}
    \tau_{\text{haze}} = 3.2\cdot 10^7 \frac{M_{\text{haze}}}{10^{-4}\,\text{M}_\odot} \left(\frac{R_{\text{Haze}}}{0.1\,\text{R}_\odot}\right)^{-2}\left(\frac{2}{\mu_e}\right),
\end{equation}
where we further assume the electron density $\mu_e\equiv \bar{Z}/\bar{A}=2$, where $\bar{Z}$ and $\bar{A}$ are the atomic number and atomic mass, respectively, corresponding to fully ionized material made of carbon, oxygen or neon. Consequently, for a characteristic radius $R_{\text{haze}}=0.1\,\text{R}_\odot$, which fills a large part of the Roche lobe, only $M_{\text{haze,crit}}=3.12\cdot 10^{-12}\,\text{M}_\odot$ is needed to significantly expand the radio-opaque region of the WD, by making it optically thick, having $\tau_{\text{haze}}\approx 1$. 

We further calculate the timescale needed to replenish this amount of material through ablation, $T_{\text{haze,form}} = M_{\text{haze,crit}}/\dot{M}_{\text{abl}}$ and the cooling timescale for the haze, which can be approximated as $T_{\text{haze,cool,init}} = GM_{\text{haze}}M_{\text{WD}}/(2L_{\text{WD}}R_{\text{WD}})$, assuming typical WD luminosities, $L_{\text{WD}}$, between $10^{-2}$ and $10^{-5}\,\text{L}
_\odot$ suitable for massive WDs older than few $100\,\text{Myr}$ \citep{2007A&A...465..249A}. We summarise these results in Table~2. Based on these estimates, we see that, among other known pulsars with massive WDs, J1928$+$1815 is unique because, within a plausible range of parameters, it cannot effectively cool and, in this way, remove the heat from the pulsar companion.

Given the suggested WD in J1928$+$1815 cannot effectively remove the heat following pulsar ablation, it must have expanded its surface layers to increase its surface area and cool more effectively \citep{1992MNRAS.254P..19S}. However, the pulsar irradiation rate of the expanded outer layers of the WD also increases. As we show in Table~2, assuming radio haze expands to the typical radii of $R_{\text{haze}}=0.1\,\text{R}_\odot$, the timescale needed to replenish the material, $T_{\text{haze, exp, keep}} = M_{\text{haze,crit}}(R_{\text{haze}})/\dot{M}_{\text{abl}}(R_{\text{haze}})$, is even shorter relative to the new cool-down timescale $T_{\text{haze,cool,exp}} = GM_{\text{haze}}M_{\text{WD}}/(2 L_{\text{WD}}R_{\text{haze}})$. Therefore, expansion acts as a runaway process, with radio haze filling the WD Roche lobe and escaping into the orbital plane, likely reaching a self-regulated state. The morphology of the eclipsing haze possibly resembles that of hot Jupiters, evaporated by close-by stellar companions and forming extended tails and circumbinary material. Therefore, because the haze may be preferably concentrated in the orbital plane, it is possible that an edge-on viewing angle improves the chance of observing extended eclipses. 

For a more visual summary of the results, we present the pulsar population from Table~1 on the $\dot{E}-P_{\text{b}}$ plane in Figure~\ref{fig_psr_cmp}. In the figure, we show the boundary between the systems that can and cannot cool effectively by solving the equation $T_{\text{haze,form}}=T_{\text{haze,cool,max}}$. Written explicitly, the expression is given by:
\begin{equation}
\dot{E}_{\text{crit}} = \frac{ 8L_{\text{WD}} a_{\text{bin}}^2 }{f R_{\text{WD}}^2} = \frac{ 8L_{\text{WD}}  (G(M_{\text{WD}}+M_{\text{NS}}))^{2/3}}{(2\pi)^{4/3} f R_{\text{WD}}^2}P_b^{4/3}
\end{equation}
Notably, the expression is given by a simple power law and does not depend on $M_{\text{haze}}$. Therefore, the condition is relatively insensitive to assumptions about opacity and radial haze density distribution. At the same time, the true boundary depends on the unknown ablation efficiency factor $f$, which we have chosen to be $0.2$, but it could also be higher, e.g., between $0.4$ and $0.6$. Likewise, the boundary depends on the mass of the WD and on its luminosity, which decreases with age. As one can see from Figure~\ref{fig_psr_cmp}, J1928$+$1815 is unique among the pulsars in Table~1 due to its inability to cool effectively. At the same time, the other known pulsars cool down faster than ablation happens, and are not expected to show large radio eclipses, consistent with observations. This distinction does not depend on the factor of several uncertainties in $\dot{E}_{\text{crit}}$.

Another possible ablation mechanism may occur if the WD is magnetic. In this case, if the magnetic field of the WD is strong enough (proportional to B$_{\rm surf}^{\!\frac{1}{3}}$), its extended magnetosphere can trap the pulsar wind material making the whole WD magnetosphere radio opaque, thus increasing the eclipse probability \citep{2000ApJ...541..335K}. This is similar to the double pulsar system J0737$-$3039, which exhibits radio eclipses due to J0737-3039B's magnetosphere 
\citep{2004ApJ...616L.131M,2005ApJ...634.1223L,2024MNRAS.534.3936L}. Despite the NS's extremely small size, a perfect edge-on alignment is not required \citep[i=$89.35^{\circ}$;][]{2021PhRvX..11d1050K}. In support of this scenario, we note that the known observed magnetic WDs tend to be more massive compared to non-magnetic WDs \citep{2015SSRv..191..111F}, similar to the massive companion in J1928$+$1815. However, no known magnetic WDs in double WD binaries are found at short orbital periods, with the only exception being when they underwent accretion in the past \citep{2021NatAs...5..648S}. Since in J1928$+$1815 and all circular systems in Table~1, the WD formed after the NS and could not have accreted from the companion, it is thus unlikely to be magnetic. Nevertheless, it is valuable to confirm this fact observationally. Should the WD indeed be magnetic, it would offer another plausible explanation for the observed eclipses. 

We conclude that the combination of He star sizes and compact orbits may indeed facilitate the occurrence of eclipses as suggested by YH2025. On the other hand: 
\begin{itemize}[noitemsep, topsep=0pt, parsep=0pt, partopsep=0pt]
    \item Our observations suggest the absence of a stripped or naked star below the minimum companion mass permitted by pulsar timing measurements.
\item Millisecond pulsars in binaries with WD companions both form more commonly and live much longer than pulsars with helium star companions. A proto-WD companion was excluded by YH2025 due to its large mass and short lifetime ($\le 10^4$\,yrs), which results in a low occurrence rate (Table~S3 of YH2025).
\item We have shown that the pulsar in J1928$+$1815, which is distinct from  all other known binary pulsars with massive WD companions, has likely significantly bloated the outer layers of its companion, producing a binary-scale haze opaque to radio waves. Therefore, it is plausible that J1928$+$1815 contains a massive WD, and was detected purely through its ability to produce eclipses, in this way introducing a natural bias for it to be close and unusually rapidly spinning.
\end{itemize}

Our current explanation for eclipsing is based on a spherically symmetric model. While YH2025 argues that a highly inclined orbital angle is necessary, we maintain that it is at least critically important. This is particularly true for systems that lie near their boundary lines in Figure~\ref{fig_psr_cmp}, where this high inclination could relax the criteria required for eclipses to occur; as a result, other massive black widow (BW) systems may exist in Table~1 but remain undetected as eclipsing binaries due to suboptimal viewing angles. Accordingly, J1928$+$1815 could be a subtype of short-period pulsars with WD companions, where significant eclipses are only observed for systems oriented within a small fraction ($\lesssim 10\%$) of possible angles. If J1928$+$1815 is indeed a massive BW system, it is unlikely to evolve into the lower mass end of the BW family. Even given that the WD would be ablated in this scenario, the timescale to evaporate the whole WD is longer than the Hubble time or the time for the system to spiral in due to GW emission. The outcome of such an inspiral will be a WD-NS merger event producing a faint SN-like transient likely resembling the observed SNe Iax \citep{2017MNRAS.467.3556B,2022MNRAS.510.3758B}. More robust brightness constraints should be obtained using the \textit{HST} due to the crowding around J1928$+$1815.

\begin{table}
\scriptsize{
\caption{J1928$+$1815 and pulsar binaries with massive WD companions}
\label{archive}
\begin{center}
\begin{tabular}{llllllllll}
\hline\hline
$\textit{Name}$ (1)&$P_0$ (2)& $P_1$ (3)& $P_b$ (4)& $A_1$ (5)& $Ecc$ (6)& $MinMass$ (7)& $Age$ (8)& $B_{surf}$ (9)& $\dot{E}$ (10)\\
& (ms)&(s/s)&(days)&(lt-s) &&($\rm M_{\odot}$)&(Gyr)&(Gauss)&(erg/s)\\
\hline 
J1928$+$1815& 10.549&3.63E-18 &0.1499&1.69  & $<$3e-5& 1.15    & 0.46   &6.3e9& 1.2e35 \\
\hline
J1141-6545& 393.898&4.30E-15 &0.1976&1.85  & 0.1718& 0.98    & 1.4e-3   &1.3e12& 2.8e33 \\
\hline
J1757-5322 &8.869& 2.63E-20  &0.4533&2.08 &e-6 &0.55  &5.34  &4.8e8&1.5e33 \\
\hline
J1802-2124 &12.647 &7.25E-20 &0.6988&3.71&e-6  &0.80  &2.76  &9.6e8&1.4e33\\
 \hline
J1525-5545 &11.359 &1.31E-19&0.9903&4.71&e-6 &0.81  &1.37  &1.2e9&3.5e33\\
 \hline
J1952+2630 &20.732 &4.27E-18 &0.3918&2.79&e-5 &0.92  &0.07  &9.5e9&1.9e34\\
 \hline
J0416+5201 &18.240 &2.32E-19 &0.3964&3.50&e-5 &1.28  &1.24  &2.0e9&1.5e33\\ 
 \hline
J0520+3722 &7.913 &3.56E-19 &0.5796&3.37&e-6 &0.85 &0.35  &1.6e9&2.8e34 \\
 \hline
J1919+1341 &11.655 &1.70E-20 &0.3703&2.34&e-5 &0.78 &10.9  &4.5e8&4.2e32\\ 
 \hline
J1943+2210 &12.870 &5.66E-20 &0.3720&2.57&e-6 &0.89  &3.61  &8.6e8&1.0e33 \\
 \hline
J1947+2304 &10.893 &4.03E-20 &0.3388&2.39&e-5 &0.87  &4.29  &6.6e8&1.2e33 \\
 \hline
J2023+2853 &11.328 &2.68E-20 &0.7182&4.00&e-5 &0.89 &6.70  &5.5e8&7.3e32 \\ 
\hline

\end{tabular}
\end{center}
The pulsar items (Pulsar name (1), Barycentric period of the pulsar (2), Time derivative of barycentric period (3), Binary period (4), Projected semi-major axis of orbit (5), Eccentricity (6), Minimum companion mass assuming i=$90^\circ$ and neutron star mass is 1.35\,$\rm M_{\odot}$ (7), Spin down age (8), Surface magnetic flux density (9) and spin down energy loss rate (10)) and values are derived from the ATNF catalog \citep{2005AJ....129.1993M}, YH2025, and \citet{2025RAA....25a4002Y}, and have been truncated for simplicity. For example, the lowercase 'e' in the eccentricity column denotes an order of magnitude estimation. A momentum of inertia of $\rm I = 1 \times 10^{45} \, \mathrm{g \, cm^{2}} $ is adopted to calculate their spin-down power \citep{2016era..book.....C}.
 \\
}
\end{table}

\begin{table}
\scriptsize{
\caption{Derived quantities for the massive black widow scenario}
\label{derived}
\begin{center}
\begin{tabular}
{lllllllllllll}
\hline\hline

$\textit{Name}$ (1)&$\textit{\.{E}}$ (2)& $\textit{R}_{\text{WD}}$ (3)& $\textit{\.{M}}_{\text{abl}}$ (4)& $\textit{T}_{\text{haze,form}}$ (5)& $\textit{T}_{\text{haze,cool,init}}$ (6)& $\textit{\.{M}}_{\text{abl,exp}}$ (7)& $\textit{T}_{\text{haze,exp,keep}}$ (8)& $\textit{T}_{\text{haze,cool,exp}}$ (9) \\
& (erg/s) & ($\mathrm{R}_\odot$) & ($\mathrm{M}_\odot$/yr) & (yr) & (yr) & ($\mathrm{M}_\odot$/yr) & (yr) & (yr) \\
\hline
J1928$+$1815 & 1.2e35 & 6.49E-03 & 4.55E-15 & 6.86E+02 & 8.69E+02 -- 8.69E-01 & 1.67E-11 & 1.87E-01 & 5.64E+01 -- 5.64E-02 \\
\hline
J1141-6545 & 2.8e33 & 8.36E-03 & 1.85E-16 & 1.69E+04 & 5.75E+02 -- 5.75E-01 & 3.17E-13 & 9.84E+00 & 4.81E+01 -- 4.81E-02 \\
\hline
J1757-5322 & 1.5e33 & 1.34E-02 & 2.70E-16 & 1.16E+04 & 2.02E+02 -- 2.02E-01 & 1.12E-13 & 2.79E+01 & 2.70E+01 -- 2.70E-02 \\
\hline
J1802-2124 & 1.4e33 & 1.03E-02 & 4.12E-17 & 7.57E+04 & 3.81E+02 -- 3.81E-01 & 3.77E-14 & 8.28E+01 & 3.92E+01 -- 3.92E-02 \\
\hline
J1525-5545 & 3.5e33 & 1.02E-02 & 6.22E-17 & 5.02E+04 & 3.90E+02 -- 3.90E-01 & 5.86E-14 & 5.32E+01 & 3.97E+01 -- 3.97E-02 \\
\hline
J1952+2630 & 1.9e34 & 9.00E-03 & 6.82E-16 & 4.57E+03 & 5.01E+02 -- 5.01E-01 & 9.35E-13 & 3.34E+00 & 4.51E+01 -- 4.51E-02 \\
\hline
J0416+5201 & 1.5e33 & 4.85E-03 & 5.58E-18 & 5.59E+05 & 1.29E+03 -- 1.29E+00 & 4.89E-14 & 6.38E+01 & 6.28E+01 -- 6.28E-02 \\
\hline
J0520+3722 & 2.8e34 & 9.76E-03 & 8.62E-16 & 3.62E+03 & 4.27E+02 -- 4.27E-01 & 9.28E-13 & 3.36E+00 & 4.17E+01 -- 4.17E-02 \\
\hline
J1919+1341 & 4.2e32 & 1.05E-02 & 3.27E-17 & 9.54E+04 & 3.63E+02 -- 3.63E-01 & 2.82E-14 & 1.11E+02 & 3.83E+01 -- 3.83E-02 \\
\hline
J1943+2210 & 1.0e33 & 9.33E-03 & 4.68E-17 & 6.67E+04 & 4.68E+02 -- 4.68E-01 & 5.76E-14 & 5.42E+01 & 4.37E+01 -- 4.37E-02 \\
\hline
J1947+2304 & 1.2e33 & 9.54E-03 & 6.90E-17 & 4.52E+04 & 4.47E+02 -- 4.47E-01 & 7.95E-14 & 3.92E+01 & 4.27E+01 -- 4.27E-02 \\
\hline
J2023+2853 & 7.3e32 & 9.33E-03 & 1.41E-17 & 2.21E+05 & 4.68E+02 -- 4.68E-01 & 1.74E-14 & 1.79E+02 & 4.37E+01 -- 4.37E-02 \\
\hline
\end{tabular}
\end{center}
Table of the derived quantities for the ablation scenario. The columns list the pulsar name (1), spin-down power (2), WD radius corresponding to companion's minimal mass (3), the pulsar ablation rate from WD surface (4), the time needed to form radio-opaque haze through pulsar ablation (5), the range of WD cooling timescales (shown as max value -- min value) (6), the ablation rate of the WD with expanded outer layers (7), the time to sustain the expanded structure through ablation (8) and the range of WD cooling timescales of the expanded structures (shown as max value -- min value) (9). We note that, assuming it is a massive WD, only J1928$+$1815 can both plausibly expand and sustain the expanded radio-haze environment.
 \\
}
\end{table}

\begin{acknowledgments}
We thank the anonymous reviewer for helpful comments and constructive suggestions. GH gratefully acknowledges Prof. Han Jinlin, PI of the FAST Galactic Plane Pulsar Snapshot survey, for sharing the pulsar details in advance. GH thanks Yang Zonglin for numerous helpful discussions. GH also appreciates the helpful discussions with Lucas Guillemot, Meng Linqi, Ylva Götberg, Li Linlin and Zhou Zhimin. This work is supported by National Natural Science Foundation of China (grant No. 12588202). A.B.~acknowledges support from the Australian Research Council (ARC) Centre of Excellence for Gravitational Wave Discovery (OzGrav), through project number CE230100016.
 
Based on observations made with the GTC telescope, in the Spanish Observatorio del Roque de los Muchachos of the Instituto de Astrofísica de Canarias, under Director’s Discretionary Time. This work has made use of data from the European Space Agency (ESA) mission {\it Gaia} (\url{https://www.cosmos.esa.int/gaia}), processed by the {\it Gaia} Data Processing and Analysis Consortium (DPAC,
\url{https://www.cosmos.esa.int/web/gaia/dpac/consortium}). Funding for the DPAC has been provided by national institutions, in particular the institutions participating in the {\it Gaia} Multilateral Agreement. This publication makes use of data products from the Two Micron All Sky Survey, which is a joint project of the University of Massachusetts and the Infrared Processing and Analysis Center/California Institute of Technology, funded by the National Aeronautics and Space Administration and the National Science Foundation.

\end{acknowledgments}

\bibliography{sample631}{}
\bibliographystyle{aasjournal}



\end{document}